\documentclass[aps,prb,twocolumn,showpacs,superscriptaddress,groupedaddress]{revtex4}
\usepackage[utf8]{inputenc}
\usepackage{graphicx}
\usepackage{epstopdf,epsfig}
\usepackage{amsmath}
\usepackage{amssymb}
\usepackage{amsbsy,amsxtra}
\usepackage{amsfonts}
\usepackage{braket}
\usepackage{dcolumn}   
\usepackage{bm}        
\usepackage{amssymb}   
\usepackage[usenames]{color}
\usepackage{natbib}
\usepackage{float}
\usepackage{gensymb}
\usepackage{hyperref}
\usepackage{subfigure}
\usepackage{lipsum}

\begin{document}

\preprint{AIP/123-QED}

\title[Sample title]{Thermodynamic conditions during growth determine the magnetic anisotropy in epitaxial thin-films of La$_{0.7}$Sr$_{0.3}$MnO$_{3}$.}

\author{J. M. Vila-Fungueiriño}
\affiliation{Center for Research in Biological Chemistry and Molecular Materials (CIQUS), University of Santiago de Compostela, 15782 Santiago de Compostela, Spain}
\author{Cong Tinh Bui}
\affiliation{Center for Research in Biological Chemistry and Molecular Materials (CIQUS), University of Santiago de Compostela, 15782 Santiago de Compostela, Spain}
\author{B. Rivas-Murias}
\affiliation{Center for Research in Biological Chemistry and Molecular Materials (CIQUS), University of Santiago de Compostela, 15782 Santiago de Compostela, Spain}
\author{E. Winkler}
\affiliation{Centro At\'{o}mico Bariloche, CNEA-CONICET, 8400 S.C.
de Bariloche, R\'{\i}o Negro, Argentina}
\author{J. Milano}
\affiliation{Centro At\'{o}mico Bariloche, CNEA-CONICET, 8400 S.C.
de Bariloche, R\'{\i}o Negro, Argentina}
\author{J. Santiso}
\affiliation{Catalan Institute of Nanoscience and Nanotechnology (ICN2), CSIC and The Barcelona Institute of Science and Technology, Campus UAB, Bellaterra, 08193 Barcelona, Spain}
\author{F. Rivadulla}
\email{f.rivadulla@usc.es}
\affiliation{Center for Research in Biological Chemistry and Molecular Materials (CIQUS), University of Santiago de Compostela, 15782 Santiago de Compostela, Spain}

\date{\today}

\begin{abstract}
The suitability of a particular material for use in magnetic devices is determined by the process of magnetization reversal/relaxation,
which in turn depends on the magnetic anisotropy. Therefore, designing new ways to control magnetic anisotropy in technologically important materials
is highly desirable. Here we show that magnetic anisotropy of epitaxial thin-films of half-metallic ferromagnet La$_{0.7}$Sr$_{0.3}$MnO$_{3}$ (LSMO) is  determined by the proximity to thermodynamic equilibrium conditions during growth.
We performed a series of X-ray diffraction and ferromagnetic resonance (FMR) experiments in two different sets of samples: the first corresponds to LSMO thin-films deposited under tensile strain on (001) SrTiO$_{3}$ by Pulsed Laser Deposition (PLD; far from thermodynamic equilibrium); the second were deposited by a slow Chemical Solution Deposition (CSD) method, under quasi-equilibrium conditions.
Thin films prepared by PLD show a in-plane cubic anisotropy with an overimposed uniaxial term. A large anisotropy constant perpendicular to the film plane was also observed in these films. However, the uniaxial anisotropy is completely suppressed in the CSD films. The out of plane anisotropy is also reduced, resulting in a much stronger in plane cubic anisotropy in the chemically synthesized films. This change is due to a different rotation pattern of MnO$_{6}$ octahedra to accomodate epitaxial strain, which depends not only on the amount of tensile stress imposed by the STO substrate, but also on the growth conditions. Our results demonstrate that the nature and magnitude of the magnetic anisotropy in LSMO can be tuned by the thermodynamic parameters during thin-film deposition.

\end{abstract}


\maketitle

\section{INTRODUCTION}

Recent developments in thin-film growth showed the enormous potential of epitaxial stress to tune the properties of thin films
at a very fundamental level. In perovskite oxides, ABO$_3$, strain accommodation occurs through a complex rotation and deformation of
corner sharing BO$_6$ octahedra\cite{PRBKoster, Borisevich, PRLSantiso}. This changes the delicate balance of bond-distances and
angles and therefore the  relative orbital occupation supporting a given magnetic or electrical interaction \cite{Heidler, NatCommGervasi,
NanolettLCO, NanolettIrene, AFMSRO, PRBAruta, NatCommZhai}. Interfacial phenomena like heterogeneous catalysis reactions
\cite{NatChemGoodenough}, and electronic reconstructions occurring at functional interfaces \cite{Pentcheva}, are also influenced by these effects.

An interesting example given its scientific relevance is the case of
half-metallic ferromagnet  La$_{0.7}$Sr$_{0.3}$MnO$_{3}$ (LSMO).
Growing epitaxial LSMO on cubic (001) SrTiO$_3$ (STO) results in an
orthorhombic (with a monoclinic distortion) unit cell of the
magnetic oxide \cite{PRBKoster, Koster2011} . Biaxial tensile stress
(a=b$>$c) imposed by the cubic substrate to the incommensurate
rhombohedral lattice of bulk LSMO induces an equal in-(out-) phase
rotation of the MnO$_6$ octahedra along the a-(b-) axis, and no
rotation along the c-axis (a$^+$a$^-$c$^0$ in the Glazer
notation\cite{Glazer}). Sandiumenge et al.\cite{PRLSantiso} proposed
a complex relaxation pattern in which several phases with different
symmetry can be distinguished in epitaxial LSMO below 25 nm. These
authors identified a critical thickness $\approx$ 2 nm for the build
up of a shear strain field, which induces a rhombohedral twined
structure and a progressive compression of the c-axis up to
$\approx$10 nm. Beyond this thickness, an elastic deformation of the
lattice without any perturbation of the octahedral tilting sets up
to $\approx$25 nm. Vailionis et al.\cite{Vailionis2014} confirmed
that the mechanism of strain relaxation changes along the film
thickness, due to a combined effect of symmetry mismatch close to the
interface, and lattice mismatch in the "bulk" of the film. They
showed that in the first $\approx$two unit cells, stress suppresses
octahedral rotations and expands the c-axis parameter; farther away
from the interface, tilting of MnO$_{6}$ octahedra reduce the
 c-axis parameter
consistent with in-plane tensile strain.

An important question is whether this complex relaxation pattern is
intrinsic to the  accommodation of biaxial tensile stress in LSMO,
or if it can be modified by growing the films under very different
conditions, thus allowing the system to explore different relaxation
paths. After all, previous studies were performed on samples
synthesized by PLD and sputtering, far from thermodynamic
equilibrium. Here we describe a comparative X-ray diffraction (XRD)
and Ferromagnetic resonance (FMR) study of LSMO thin films deposited
on STO by Pulsed Laser Deposition (PLD; far from thermodynamic
equilibrium), and by a slow chemical deposition method (CSD; close
to thermodynamic equilibrium). Our results demonstrate that 
magnetic anisotropy in LSMO depends on the conditions during film
growth. As a result, a different patter of rotation of MnO$_{6}$ octahedra is accomodated in epitaxial films 
synthesized by CSD.

\section{EXPERIMENTAL DETAILS}

Thin films of LSMO of different thicknesses were grown on (001)
TiO$_2$-terminated SrTiO$_3$ (STO) substrates. For the PLD filmms we used an
excimer laser (F-Kr, 248 nm) operating at 5 Hz and a fluence of 0.8
J/cm$^2$. The films were deposited at 800 ºC and 200 mTorr of
O$_2$. For chemically grown films (CSD), a precursor solution was
spun-coated on similar substrates, and annealed at high temperature,
as described in \cite{ACSJose2015}. The thickness of both type of
films was determined by X-ray reflectivity and TEM analysis of
cross-section lamellae. 
X-band($\omega/2\pi\sim$9.4 GHz) FMR experiments were performed in a
Bruker-X spectrometer at different temperatures, with the magnetic
field applied rotating paralell to the film-plane.

\section{RESULTS AND DISCUSSION}

\begin{figure}
\includegraphics[width=3.5 in]{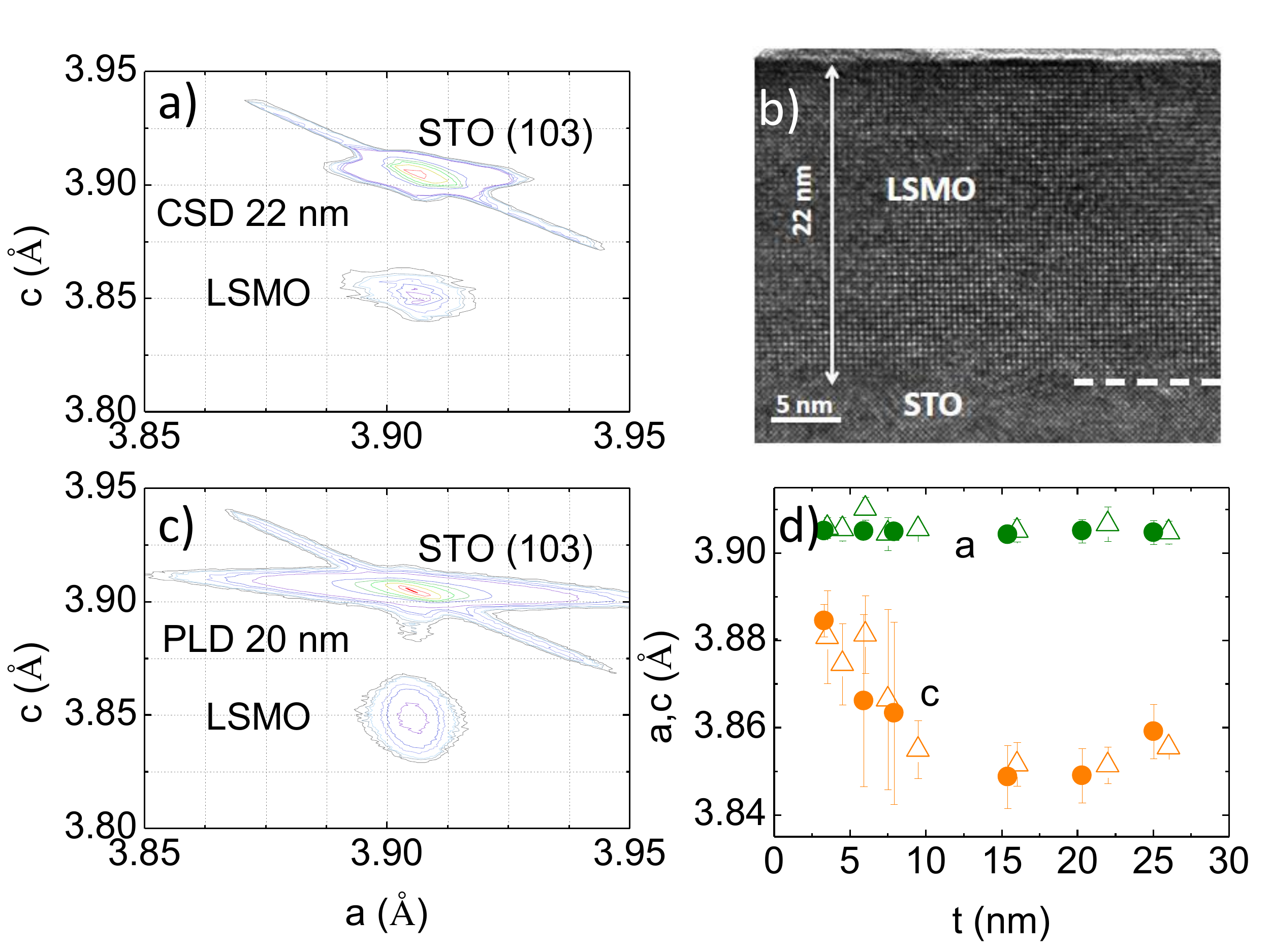}
    \caption{X-ray maps around the (103) Bragg reflection of the perovskite for two LSMO thin-films deposited on STO by  CSD (a) and PLD (c). TEM image of a cross-section lamella of a LSMO thin-film synthesized by CSD (b). The evolution of lattice parameters for several samples of different thicknesses are shown in (d).}
\label{FigXRD1}
\end{figure}

In Figure \ref{FigXRD1} we show  a summary of structural results representative of the quality of the samples
studied in this work. X-ray reciprocal space maps (RSM) around the (103) reflection of the perovskite for $\approx$ 20 nm
thick films show that they grow with in-plane lattice parameters well
matched to the STO substrate, and without evidence of lattice
relaxation (Figure \ref{FigXRD1}a), c)). This is true for every film studied in this work, irrespective of the thickness or the deposition method (CSD or PLD). 
A high-resolution cross-section TEM image of a thin film of LSMO synthesized by CSD is shown in Figure \ref{FigXRD1}b). The image is representative of the good crystalline quality and abrupt interfaces of all the CSD films reported in this work. 

\begin{figure}
\includegraphics[width=3.5 in]{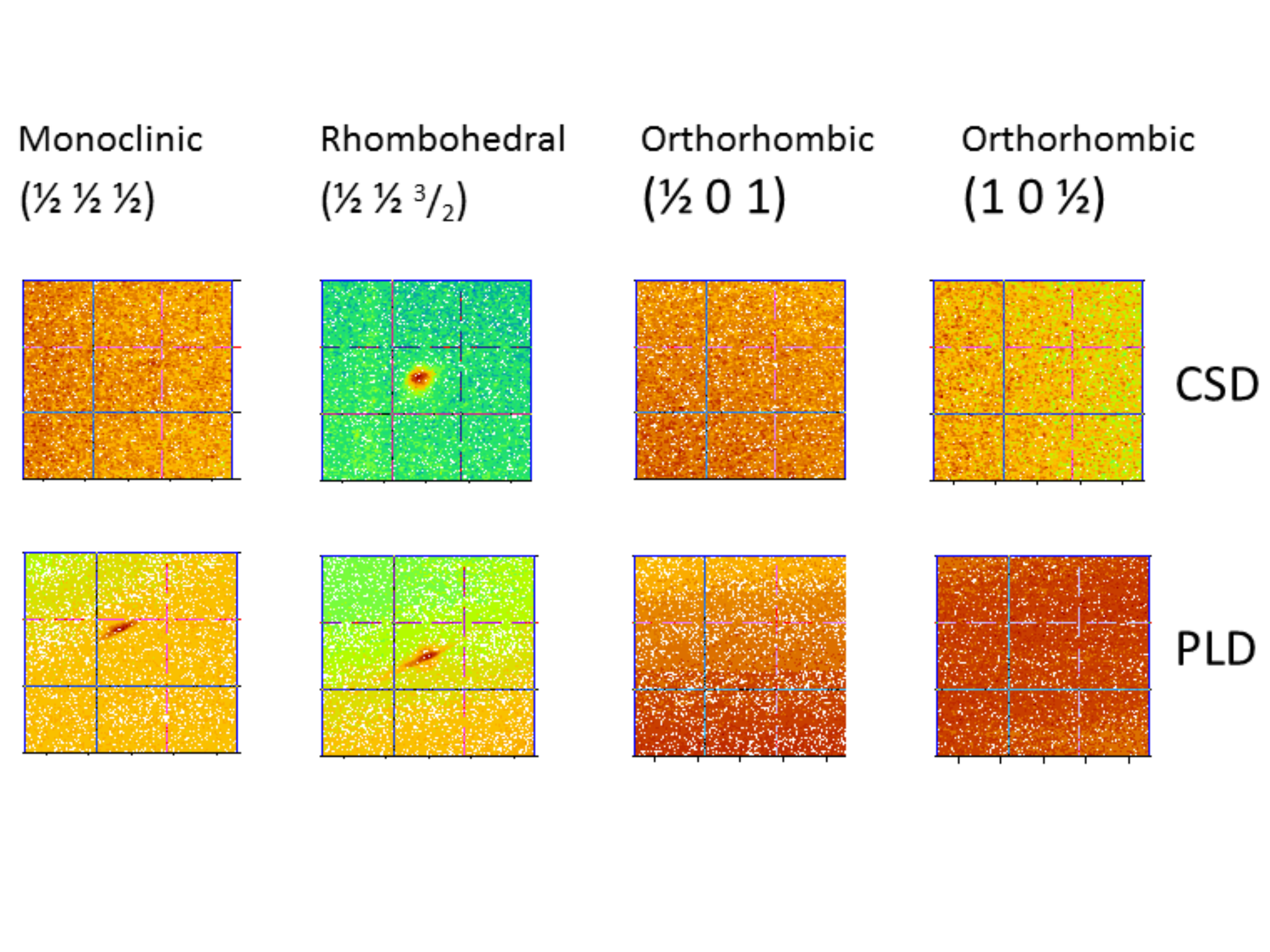}
    \caption{Half-order Bragg reflections for thin films prepared by CSD (22 nm, top) and PLD (20 nm, bottom) respectively. Different reflections characteristic of different crystallographic phases are analyzed.}
\label{FigXRD2}
\end{figure}

\begin{figure}
\includegraphics[width=3.5 in]{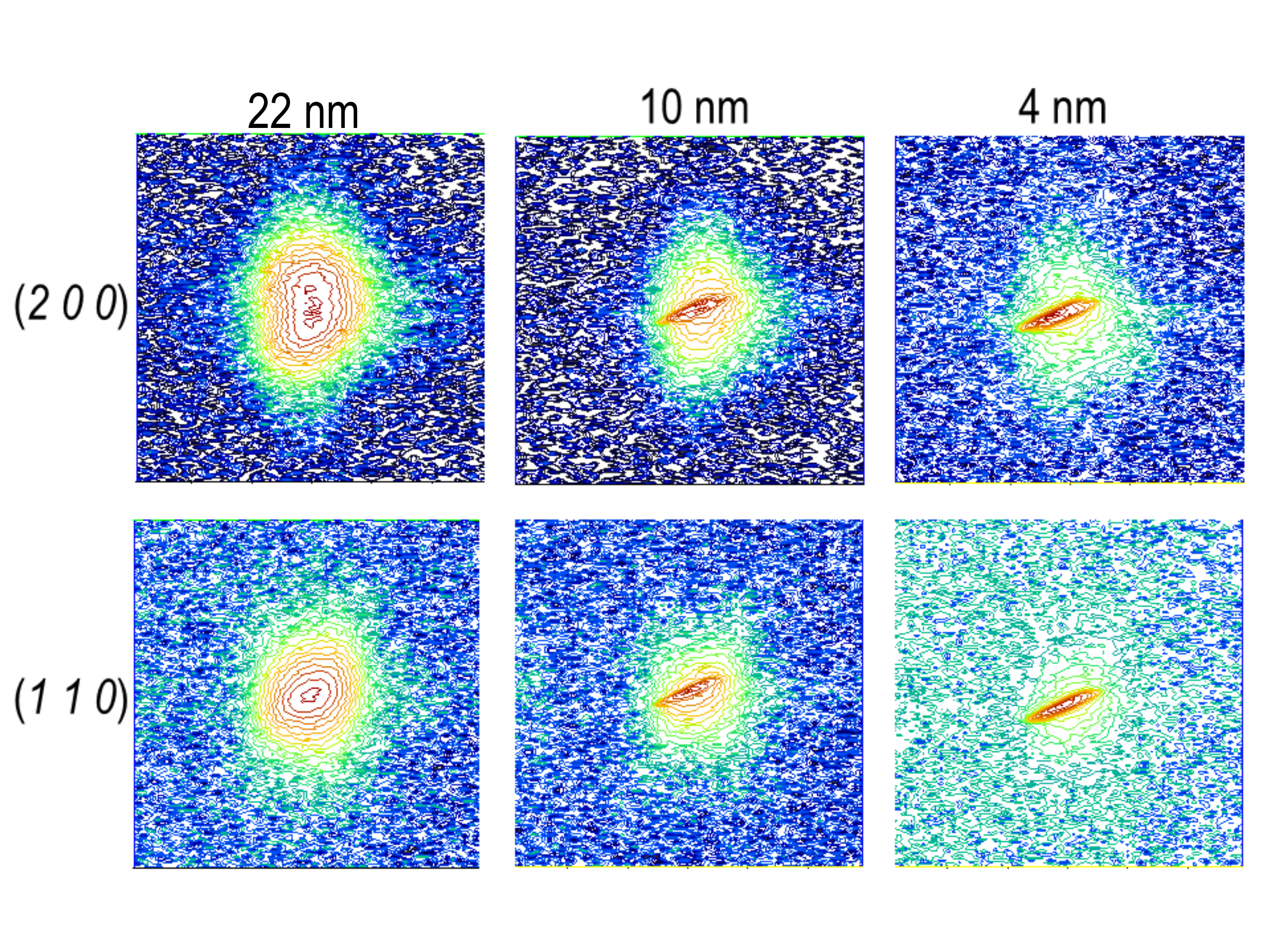}
    \caption{In plane (200) and (110) Bragg reflections for thin films of LSMO prepared by CSD, with different thickness.}
\label{FigXRD3}
\end{figure}

The dependence of the lattice parameters on the thickness is shown in Figure \ref{FigXRD1}d). The c-axis length shows a non-monotonic
dependence with the film thickness, passing through a minimum between 10 and 15 nm. Similar behavior was previously reported by Sandiumenge et
al.\cite{PRLSantiso} for LSMO films synthesized by rf-sputtering. These authors suggested that $t_{m}$ marks a crossover from a monoclinic to a
homogeneously strained rhombohedral phase. Our results show that the existence of this minimum occurs for PLD and CSD samples, therefore suggesting a universal relaxation mechanism depending only on the total strain imposed by the substrate.
However, the incommensurability of rhombohedral LSMO to the cubic (001) surface of STO has been suggeseted to result in an orthorhombic symmetry
with a monoclinic distortion (P2$_1$/\textit{m})\cite{Koster2011}.
In order to identify the crystal structure in our films, we have performed a careful XRD analysis around different half order reflections. (H/2, K/2, L/2) reflections are characteristic of a monoclinic or triclinic symmetry, with H=K=L being an extinction for the rhombohedral R-3C group. 
For the 20 nm thick sample  prepared by PLD, we observed a clear signal around the (1/2,1/2, 1/2) and (1/2, 1/2, 3/2) reflections, as shown in Figure \ref{FigXRD2}. These are consistent with a rhombohedral (R-3c) phase, with a monoclinic distortion. Although the (1/2,0,1) and (1,0,1/2) reflections have not been observed in our films, a orthorhombic phase cannot be completely discarded due to the small intensity characteristic of these reflections, particularly in thin-films.
On the other hand, half order reflections at L=3/2 and absence at L=1/2 in films prepared by CSD are consistent with a dominant rhombohedral (R-3c) phase. 
Also, from the analysis of in-plane (200) and (110) peaks, a fully structural coherence with the substrate is observed along the whole thickness of the films prepared by CSD (see Figure \ref{FigXRD3}). No satellites peaks or diffuse scattering associated to twinnigs or strong mosaicity are observed in these samples.

Given the equal in-plane tensile stress impossed by the substrate along the a/b directions, these structural results therefore suggests a different rotation pattern of the MnO$_6$ octahedra to accommodate the tensile stress in samples synthesized by CSD with respect to PLD.
We want to remark that this result is reproducible in different samples prepared from CSD under similar conditions.

\begin{figure}
\includegraphics[width=3.5 in]{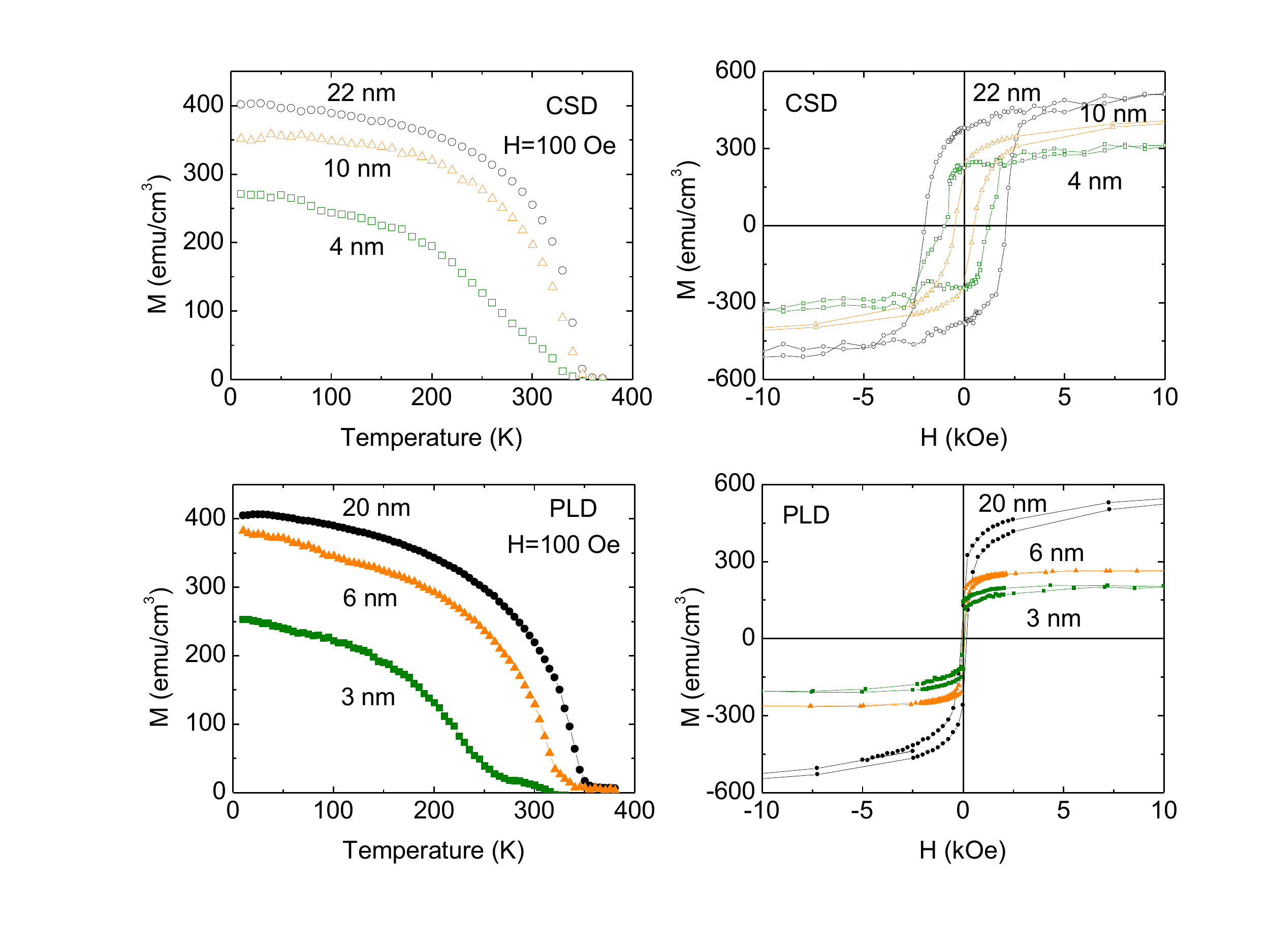}
    \caption{Temperature and field dependence (at 10 K) of the magnetization for CSD (top) and PLD (bottom) thin films.
    The hysteresis loops were measured at 10 K with the magnetic field in the film plane along the (100) axis of STO. Saturation of the magnetization occurs
    at H$<$1.5 T.}
\label{FigMvsTH}
\end{figure}

The structural difference reported in Figure \ref{FigXRD2} is also manifested in the magnetic properties of the CSD and PLD films (see Figure \ref{FigMvsTH}).
The magnetic moment at saturation and the Curie temperature (T$_c$) of $\approx$ 20 nm thick films are close to the bulk values (590
emu/cm$^3$ and 350 K) and are very similarity in both sets of samples, discarding any significant variation in their stoichiometry.
However, the coercive field (H$_c$) shows a completely different behavior: while the samples synthesized by PLD show a very small,
bulk-like, H$_c \approx$ 50-100 Oe, it increases by an order of magnitude in the films synthesized by CSD.  Therefore, the change in H$_c$ probably implies different magnetocrystalline  anisotropy between the CSD and PLD films. This could be due to differences in the strength of Mn-O-Mn exchange interactions along different directions of the crystal, as a result of the structural differences identified before.
However, to give a definitive proof of this subtle structural distortion in thin-films prepared by different methods is very challenging using conventional laboratory XRD equipment.

To avoid this difficulty, the evolution of the structural parameters with thickness was studied indirectly by ferromagnetic resonance.
FMR is a technique very sensitive to small variations in the magnitude of the different magnetic anisotropy terms. Different
rotation patterns of the MnO$_6$ octahedra along different directions of the crystal will change the orbital overlap and,
through spin-orbit coupling also change the magnetocrystalline anisotropy of the films. We will show that these changes measured by FMR can be
correlated with the structural distortions in the films.

\begin{figure}
\includegraphics[width=3.5 in]{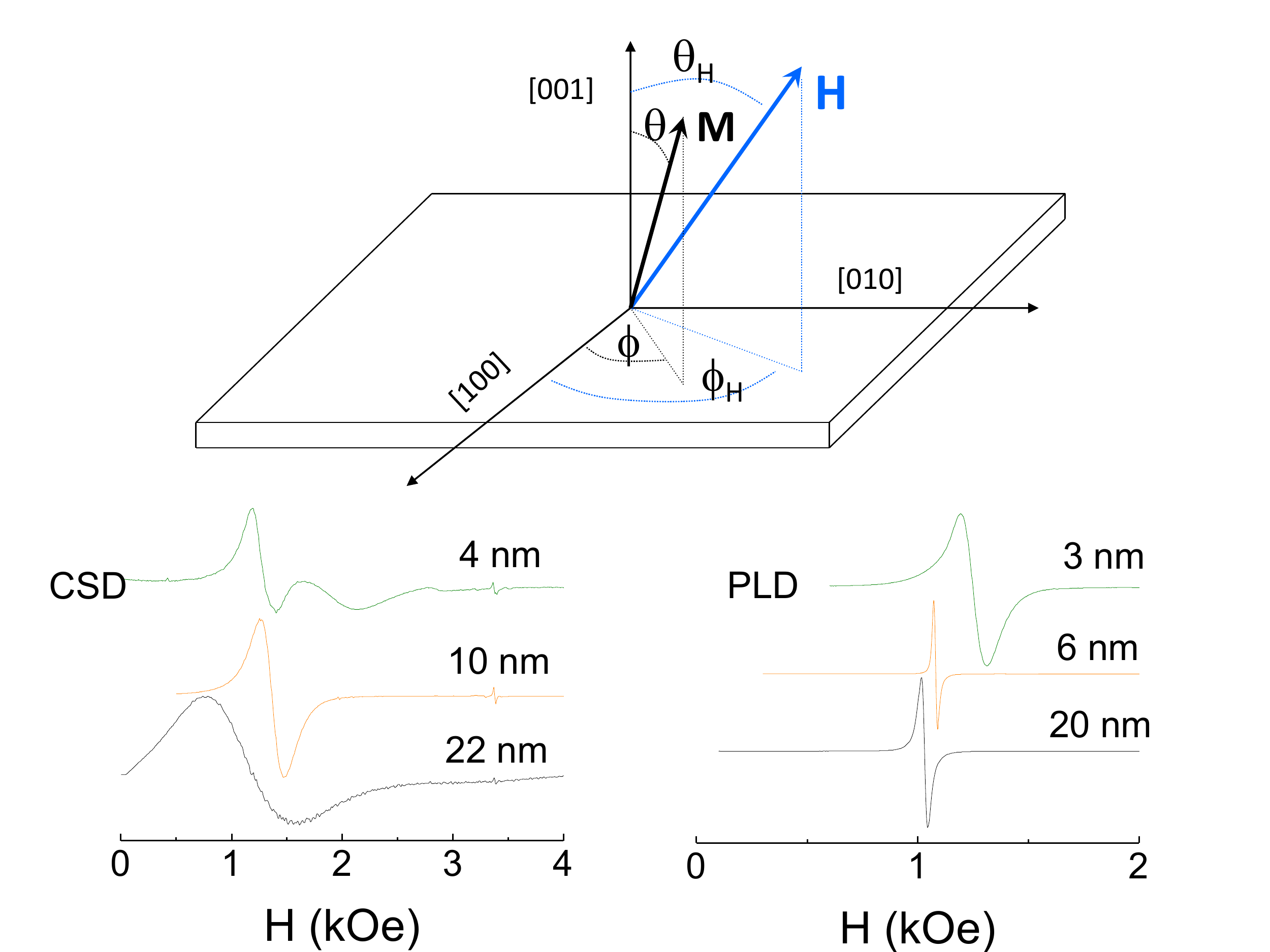}
    \caption{Top: Scheme of the coordinate system used in
    Eq.\ref{SmitBeljers1} and \ref{energy} related to the crystal axis of the STO substrate.
    Bottom: FMR lines for thin-films of LSMO of different thickness, synthesized by CSD (left) and PLD (right).
    The experiments were performed at 200 K.}
\label{FigFMRSchemeandLines}
\end{figure}

The angular dependence of the resonance field ($H_{r}$) in a FMR experiment can be evaluated at the magnetization equilibrium angles,
$\theta_{0}$ and $\phi_{0}$ for the different orientation of the magnetic field \cite{SmitBeljers}:

\begin{equation}\label{SmitBeljers1}
 \Big( \frac{\omega}{\gamma} \Big)^{2}=\frac{1}{M^{2} \sin^{2}\theta}
\bigg[\frac{\partial ^{2} F}{\partial \theta ^{2}} \frac{\partial^{2}
F}{\partial\phi ^{2}} - \Big( \frac{\partial^{2} F}{\partial \theta \partial
\phi} \Big) ^{2} \bigg]_{\theta_{0},\phi_{0}}
\end{equation}

\noindent where $\omega$ is the angular frequency, M is the saturation magnetization, $F$ is the free energy of the system
and $\gamma= g \mu_{B}/\hbar$, where $g$ is the gyromagnetic factor and $\mu_{B}$ is the Bohr magneton. Based on the structural results,
three different anisotropy terms were included in the free energy expression: a biaxial in plane anisotropy constant $K_{4}^{IP}$, an
in plane uniaxial anisotropy constat $K_{u}$, and a perpendicular out of plane anisotropy constant $K_{out}$ along [001]\cite{MilanoPRB92}:

\begin{equation}
\begin{aligned} \label{energy}
F =  - \mu_{0} \textbf{H}. \textbf{M} + \frac{\mu_{0}}{2} M^{2}
\cos^{2}\theta - \frac{K_{4}^{IP}}{4} \sin^{4}\theta \sin ^{2} 2 \phi \\
- K_{out} \cos^{2}\theta + K_{u} \sin^{2} \theta \cos^{2}
(\phi-\frac{\pi}{4})
\end{aligned}
\end{equation}

where the first and second terms correspond to the Zeeman and demagnetization energy, respectively. The vacuum permeability is
given by $\mu_{0}$, and $\theta$ and $\phi$ are the polar and azimuthal angles of the magnetization vector, according to the
scheme of Figure \ref{FigFMRSchemeandLines}. In this way the values of $K_{4}^{IP}$, $K_{u}$, and $K_{out}$ can be determined from
fittings of experimental $H_{r}$($\theta$,$\phi$) curves by solving self-consistently Eq. (1) and (2). \cite{Vittoria}

Following the formalism explained above, the thickness dependence of the magnetocrystalline anisotropy in our films were obtained from
FMR experiments with H rotating in the plane of the films, i.e. $\phi$$_H$=0-360º, $\theta$$_H$=90º. The experiments were
performed at 200 K, except for the thinner samples, which were taken at 150 K to ensure that the samples are completely magnetized at the
resonance field. The FMR spectra show a single Lorentzian line in all cases, except for thinner CSD films (see Figure \ref{FigFMRSchemeandLines}).  In this case two broad overimposed lines precludes the accurate analysis of their resonance field, so we excluded these samples from the discussion.

\begin{figure}
\includegraphics[width=3.5 in]{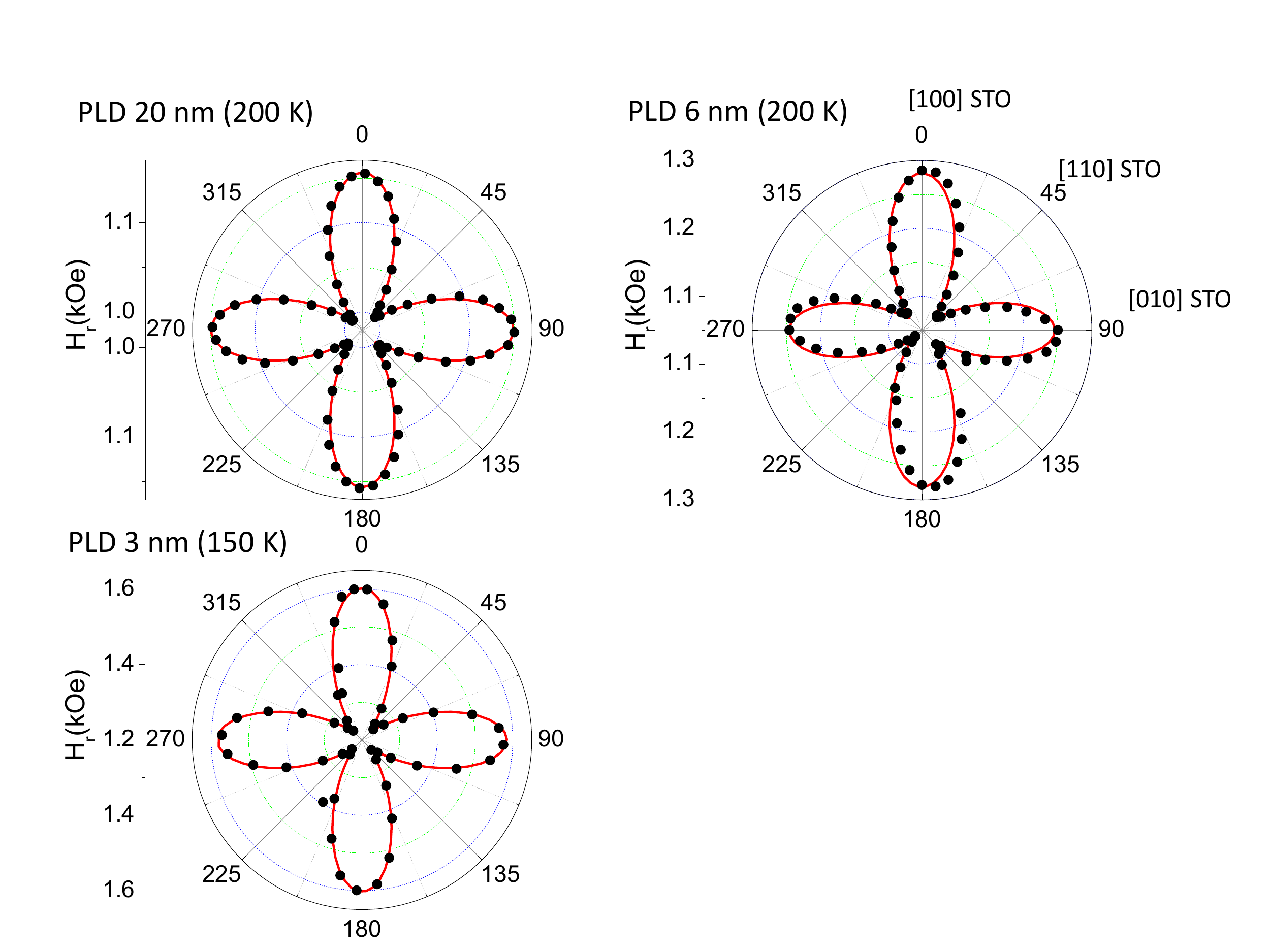}
    \caption{In-plane angular dependence ($\phi$$_H$=0-360º, $\theta$$_H$=90º) of the PLD films resonance field, H$_r$. The spectra were acquired at 200 K, except for the thinnest sample that was taken at 150 K to ensure a good magnetization of the sample. The continuous lines correspond to the best fit numerically obtained from Eq. \ref{SmitBeljers1} and \ref{energy} (see text).}
\label{FigFMRPLD}
\end{figure}

\begin{figure}
\includegraphics[width=3.5 in]{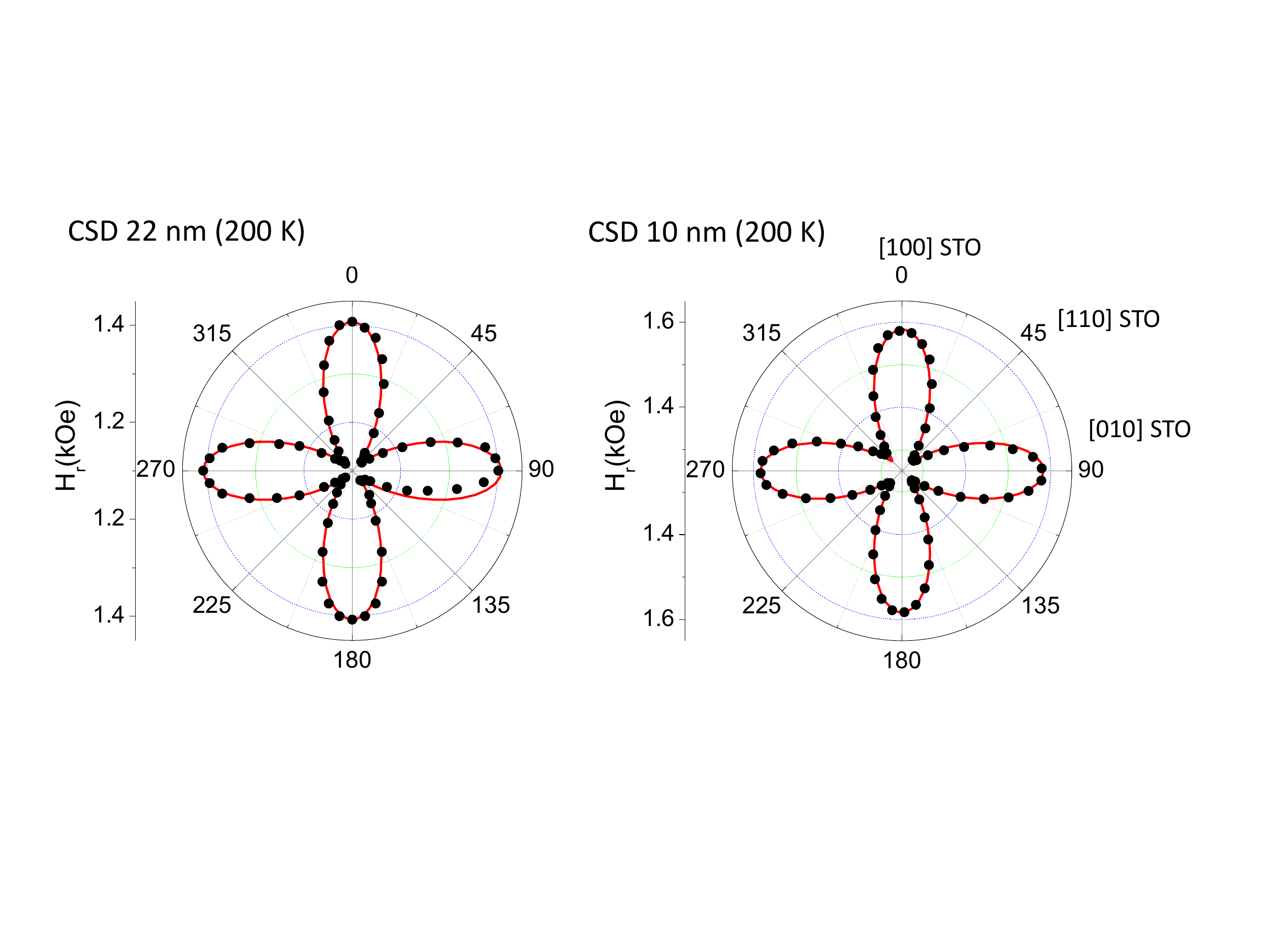}
    \caption{In-plane angular dependence ($\phi$$_H$=0-360º, $\theta$$_H$=90º) of the CSD films resonance field, H$_r$. The spectra were acquired at 200 K. The continuous lines correspond to the best fit numerically obtained from Eq. \ref{SmitBeljers1} and \ref{energy} (see text).}
\label{FigFMRCSD}
\end{figure}

The angular dependence of H$_r$($\phi$$_H$, $\theta$$_H$=90º) is shown in Figures \ref{FigFMRPLD} and \ref{FigFMRCSD}. All samples
show a clear biaxial anisotropy with the easy axis along the $<110>$ direction of STO (the diagonal directions of the (001) substrate),
and the hard axis coinciding with the $<100>$ directions of STO (the sides of the substrates), which is in agreement with previous
reports \cite{Steenbeck,EomAPL}. Note that contrary to magnetization, in a FMR experiment the maximum and minimum of $H_{r}$ mark the
hard and easy magnetization axis directions, respectively. The fitting to Eq. (1) and (2) is also shown as continuous
lines over the experimental data. To improve the accuracy of the fitting, the values of the saturation magnetization were obtained from the
experimental M(H) measurements in each sample, at the same temperature as the FMR experiments. Also the $g-$factor was set to
$g=2.0$, as generally observed in bulk La$_{0.7}$Sr$_{0.3}$MnO$_{3}$ \cite{Alejandro, Ivanshin}.

The anisotropy constants obtained from the fittings are listed in Table \ref{table1}. All PLD films, irrespective of their thickness, are characterized by a biaxial anisotropy constant $K_{4}^{IP} \approx$2 kJ/m$^3$, plus an order of magnitude smaller uniaxial anisotropy $K_{u}$. The existence of these two anisotropy
terms was previously reported from magnetization measurements in LSMO under tensile strain by several authors\cite{PRBKoster,Boschker2011,Steenbeck,BelmeguenaiPRB2010}.
The monoclinically distorted unit cell of LSMO results from a different rotation pattern of the MnO$_6$ octahedra along the $<110>$ axis (a$^+$a$^-$c$^0$). This produces an important difference in the magnitude of the orbital overlap along the equivalent $<110>$ easy axis directions, introducing the extra uniaxial anisotropy term. We also identified fro this analysis an important out of plane anisotropy constant $K_{out}$. This term may have its origin in the contribution of different factors, like magnetocrystalline anisotropy \cite{cullity, MilanoPRB92}, interface effects \cite{Ikeda2010}, domain shape \cite{Vidal2012}, or most probably in this case, interfacial stress \cite{Butera2010}. Although this effect is normally neglected, we show here that it can be an appreciable contribution in epitaxially stressed films. 

\begin{table}
\caption{\label{table1} Anisotropy energies obtained from the fits
of the FMR curves and the corresponding saturation magnetization for
the PLD and CSD films. }

\begin{tabular}{ccccccccccc}
 &t &K$_{ip}^4$ & K$_u$ & K$_{out}$ & $M$ \\
 & nm & kJ/m$^3$ & kJ/m$^3$ & kJ/m$^3$ & emu/cm$^3$ \\
\hline
PLD \\
&  3 &  2.0(2) & 0.10(1) & 30(3) & 200 \\
&6&  1.3(1) & 0.25(3) & 60(6) & 260 \\
&20& 2.3(2) & 0.10(1) & 40(4) & 580 \\
CSD \\
&10&  3.5(3) & 0 & 10(1) & 440 \\
&25&  4.5(5) & 0 & 5.0(5) & 590 \\
\end{tabular}
\end{table}

On the other hand, the situation is completely different in the films synthesized by CSD: the value of $K_{4}^{IP}$ is much larger in these films compared to PLD, and most important, the uniaxial anisotropy term vanishes, $K_u\approx$0. The out of plane anisotropy constant $K_{out}$ is also
much smaller than in the PLD films, reflecting a different contribution from interfacial epitaxial stress.
A larger in-plane cubic anisotropy is in qualitative agreement with larger in-plane coercivity, as observed in Figure \ref{FigMvsTH}. However, the estimated H$_C \sim2K^{4} _{ip}/M$ for these films is in the 80-200 Oe range, smaller than observed, which calls for further relevant effect of magnetic inhomogeneities 
which could act as pinning centers for domain walls\cite{Kronmuller}. This is also consistent with the much wider FMR lines observed in CSD with respect to PLD films.

Following the argument before, the absence of $K_u$ in CSD films indicate a similar orbital overlap along the $<110>$ axis, which in turn requires an equivalent rotation of the MnO$_6$ octahedra along the a and b axis. The most plausible possibility is a$^+$a$^+$c$^0$ (Glazer tilt system number 16),
compatible with unit cell parameters a=b$>$c under tensile stress (tetragonal, space group $I4/mmm$) and our previous X-ray analysis (see Figure \ref{FigXRD2}). The results of CSD are then consistent with an elastic deformation of LSMO, without any significant anisotropy in the octahedral tilting
along the direction of the easy axis.

\begin{figure}
\includegraphics[width=3.5 in]{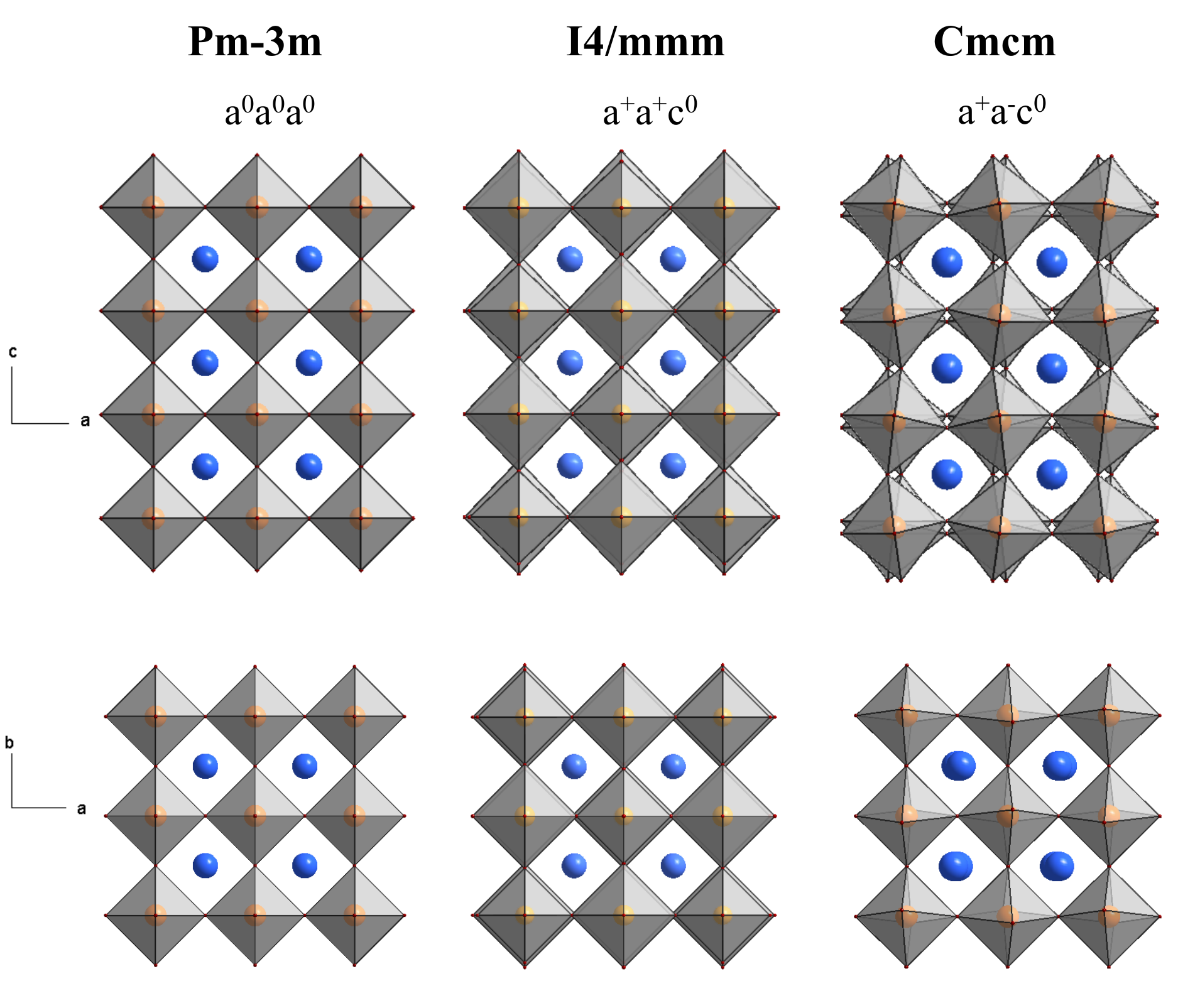}
    \caption{Rotation patern of the MnO$_{6}$ octahedra for the space groups adopted by the epitaxial LSMO thin-films synthesized by PLD ($Cmcm$) and CSD ($I4/mmm$). The situation for cubic STO is also included ($Pm-3m$), for comparison.}
\label{FigOcthedralTilting}
\end{figure}

The arrangements of MnO$_6$ octahedra compatible with the analysis of the experimental FMR and XRD in epitaxial films of LSMO synthesized by PLD and CSD are shown in Figure \ref{FigOcthedralTilting}.

In summary, we have demonstrated that the characteristic in-plane uniaxial component of the magnetic anisotropy can be completely
suppressed in chemically prepared thin-films of LSMO. This implies different mechanisms of octahedral rotation to accommodate the
biaxial tensile stress, depending on the growth conditions. Magnetic anisotropy determines the switching and relaxation of magnetization,
and therefore the results presented here are not only interesting from a fundamental point of view, but they must be considered for
applications of half-metallic ferromagnet LSMO in different types of devices. Finally, we would like to remark the enormous possibilities
offered by the sensitivity of FMR for the indirect study of subtle structural changes in ultrathin films, using conventional laboratory
equipment.

\begin{acknowledgments}
This work was supported by the European Research Council (ERC StG-259082, 2DTHERMS), Xunta de Galicia (2012-Projet No. CP072) and by the Ministry of Science of Spain (Project No. MAT2013-44673-R). J. M. V. F. also acknowledges the same organization for an FPI grant. E.W. and J.M. thank UNCuyo Argentina for Grant No C011.
\end{acknowledgments}


\begin{thebibliography}{30}
\expandafter\ifx\csname natexlab\endcsname\relax\def\natexlab#1{#1}\fi
\expandafter\ifx\csname bibnamefont\endcsname\relax
  \def\bibnamefont#1{#1}\fi
\expandafter\ifx\csname bibfnamefont\endcsname\relax
  \def\bibfnamefont#1{#1}\fi
\expandafter\ifx\csname citenamefont\endcsname\relax
  \def\citenamefont#1{#1}\fi
\expandafter\ifx\csname url\endcsname\relax
  \def\url#1{\texttt{#1}}\fi
\expandafter\ifx\csname urlprefix\endcsname\relax\def\urlprefix{URL }\fi
\providecommand{\bibinfo}[2]{#2}
\providecommand{\eprint}[2][]{\url{#2}}

\bibitem[{\citenamefont{Vailionis et~al.}(2011)\citenamefont{Vailionis,
  Boschker, Siemons, Houwman, Blank, Rijnders, and Koster}}]{PRBKoster}
\bibinfo{author}{\bibfnamefont{A.}~\bibnamefont{Vailionis}},
  \bibinfo{author}{\bibfnamefont{H.}~\bibnamefont{Boschker}},
  \bibinfo{author}{\bibfnamefont{W.}~\bibnamefont{Siemons}},
  \bibinfo{author}{\bibfnamefont{E.~P.} \bibnamefont{Houwman}},
  \bibinfo{author}{\bibfnamefont{D.~H.~A.} \bibnamefont{Blank}},
  \bibinfo{author}{\bibfnamefont{G.}~\bibnamefont{Rijnders}}, \bibnamefont{and}
  \bibinfo{author}{\bibfnamefont{G.}~\bibnamefont{Koster}},
  \bibinfo{journal}{Phys. Rev. B} \textbf{\bibinfo{volume}{83}},
  \bibinfo{pages}{064101} (\bibinfo{year}{2011}),
  \urlprefix\url{http://link.aps.org/doi/10.1103/PhysRevB.83.064101}.

\bibitem[{\citenamefont{Borisevich et~al.}(2010)\citenamefont{Borisevich,
  Chang, Huijben, Oxley, Okamoto, Niranjan, Burton, Tsymbal, Chu, Yu
  et~al.}}]{Borisevich}
\bibinfo{author}{\bibfnamefont{A.~Y.} \bibnamefont{Borisevich}},
  \bibinfo{author}{\bibfnamefont{H.~J.} \bibnamefont{Chang}},
  \bibinfo{author}{\bibfnamefont{M.}~\bibnamefont{Huijben}},
  \bibinfo{author}{\bibfnamefont{M.~P.} \bibnamefont{Oxley}},
  \bibinfo{author}{\bibfnamefont{S.}~\bibnamefont{Okamoto}},
  \bibinfo{author}{\bibfnamefont{M.~K.} \bibnamefont{Niranjan}},
  \bibinfo{author}{\bibfnamefont{J.~D.} \bibnamefont{Burton}},
  \bibinfo{author}{\bibfnamefont{E.~Y.} \bibnamefont{Tsymbal}},
  \bibinfo{author}{\bibfnamefont{Y.~H.} \bibnamefont{Chu}},
  \bibinfo{author}{\bibfnamefont{P.}~\bibnamefont{Yu}}, \bibnamefont{et~al.},
  \bibinfo{journal}{Phys. Rev. Lett.} \textbf{\bibinfo{volume}{105}},
  \bibinfo{pages}{087204} (\bibinfo{year}{2010}),
  \urlprefix\url{http://link.aps.org/doi/10.1103/PhysRevLett.105.087204}.

\bibitem[{\citenamefont{Sandiumenge et~al.}(2013)\citenamefont{Sandiumenge,
  Santiso, Balcells, Konstantinovic, Roqueta, Pomar, Espinós, and
  Martínez}}]{PRLSantiso}
\bibinfo{author}{\bibfnamefont{F.}~\bibnamefont{Sandiumenge}},
  \bibinfo{author}{\bibfnamefont{J.}~\bibnamefont{Santiso}},
  \bibinfo{author}{\bibfnamefont{L.}~\bibnamefont{Balcells}},
  \bibinfo{author}{\bibfnamefont{Z.}~\bibnamefont{Konstantinovic}},
  \bibinfo{author}{\bibfnamefont{J.}~\bibnamefont{Roqueta}},
  \bibinfo{author}{\bibfnamefont{A.}~\bibnamefont{Pomar}},
  \bibinfo{author}{\bibfnamefont{J.~P.} \bibnamefont{Espinós}},
  \bibnamefont{and}
  \bibinfo{author}{\bibfnamefont{B.}~\bibnamefont{Martínez}},
  \bibinfo{journal}{Phys. Rev. Lett.} \textbf{\bibinfo{volume}{110}},
  \bibinfo{pages}{107206} (\bibinfo{year}{2013}),
  \urlprefix\url{http://link.aps.org/doi/10.1103/PhysRevLett.110.107206}.

\bibitem[{\citenamefont{Heidler et~al.}(2015)\citenamefont{Heidler, Piamonteze,
  Chopdekar, Uribe-Laverde, Alberca, Buzzi, Uldry, Delley, Bernhard, and
  Nolting}}]{Heidler}
\bibinfo{author}{\bibfnamefont{J.}~\bibnamefont{Heidler}},
  \bibinfo{author}{\bibfnamefont{C.}~\bibnamefont{Piamonteze}},
  \bibinfo{author}{\bibfnamefont{R.~V.} \bibnamefont{Chopdekar}},
  \bibinfo{author}{\bibfnamefont{M.~A.} \bibnamefont{Uribe-Laverde}},
  \bibinfo{author}{\bibfnamefont{A.}~\bibnamefont{Alberca}},
  \bibinfo{author}{\bibfnamefont{M.}~\bibnamefont{Buzzi}},
  \bibinfo{author}{\bibfnamefont{A.}~\bibnamefont{Uldry}},
  \bibinfo{author}{\bibfnamefont{B.}~\bibnamefont{Delley}},
  \bibinfo{author}{\bibfnamefont{C.}~\bibnamefont{Bernhard}}, \bibnamefont{and}
  \bibinfo{author}{\bibfnamefont{F.}~\bibnamefont{Nolting}},
  \bibinfo{journal}{Phys. Rev. B} \textbf{\bibinfo{volume}{91}},
  \bibinfo{pages}{024406} (\bibinfo{year}{2015}),
  \urlprefix\url{http://link.aps.org/doi/10.1103/PhysRevB.91.024406}.

\bibitem[{\citenamefont{Pesquera et~al.}(2012)\citenamefont{Pesquera, Herranz,
  Barla, Pellegrin, Bondino, Magnano, Sánchez, and
  Fontcuberta}}]{NatCommGervasi}
\bibinfo{author}{\bibfnamefont{D.}~\bibnamefont{Pesquera}},
  \bibinfo{author}{\bibfnamefont{G.}~\bibnamefont{Herranz}},
  \bibinfo{author}{\bibfnamefont{A.}~\bibnamefont{Barla}},
  \bibinfo{author}{\bibfnamefont{E.}~\bibnamefont{Pellegrin}},
  \bibinfo{author}{\bibfnamefont{F.}~\bibnamefont{Bondino}},
  \bibinfo{author}{\bibfnamefont{E.}~\bibnamefont{Magnano}},
  \bibinfo{author}{\bibfnamefont{F.}~\bibnamefont{Sánchez}}, \bibnamefont{and}
  \bibinfo{author}{\bibfnamefont{J.}~\bibnamefont{Fontcuberta}},
  \bibinfo{journal}{Nature Communications} \textbf{\bibinfo{volume}{3}},
  \bibinfo{pages}{1189} (\bibinfo{year}{2012}).

\bibitem[{\citenamefont{Qiao et~al.}(2015)\citenamefont{Qiao, Jang, Singh, Gai,
  Xiao, Mehta, Vasudevan, Tselev, Feng, Zhou et~al.}}]{NanolettLCO}
\bibinfo{author}{\bibfnamefont{L.}~\bibnamefont{Qiao}},
  \bibinfo{author}{\bibfnamefont{J.~H.} \bibnamefont{Jang}},
  \bibinfo{author}{\bibfnamefont{D.~J.} \bibnamefont{Singh}},
  \bibinfo{author}{\bibfnamefont{Z.}~\bibnamefont{Gai}},
  \bibinfo{author}{\bibfnamefont{H.}~\bibnamefont{Xiao}},
  \bibinfo{author}{\bibfnamefont{A.}~\bibnamefont{Mehta}},
  \bibinfo{author}{\bibfnamefont{R.~K.} \bibnamefont{Vasudevan}},
  \bibinfo{author}{\bibfnamefont{A.}~\bibnamefont{Tselev}},
  \bibinfo{author}{\bibfnamefont{Z.}~\bibnamefont{Feng}},
  \bibinfo{author}{\bibfnamefont{H.}~\bibnamefont{Zhou}}, \bibnamefont{et~al.},
  \bibinfo{journal}{Nano Letters} \textbf{\bibinfo{volume}{15}},
  \bibinfo{pages}{4677} (\bibinfo{year}{2015}).

\bibitem[{\citenamefont{Marín et~al.}(2015)\citenamefont{Marín, Rodríguez,
  Magén, Snoeck, Arras, Lucas, Morellón, Algarabel, De~Teresa, and
  Ibarra}}]{NanolettIrene}
\bibinfo{author}{\bibfnamefont{L.}~\bibnamefont{Marín}},
  \bibinfo{author}{\bibfnamefont{L.~A.} \bibnamefont{Rodríguez}},
  \bibinfo{author}{\bibfnamefont{C.}~\bibnamefont{Magén}},
  \bibinfo{author}{\bibfnamefont{E.}~\bibnamefont{Snoeck}},
  \bibinfo{author}{\bibfnamefont{R.}~\bibnamefont{Arras}},
  \bibinfo{author}{\bibfnamefont{I.}~\bibnamefont{Lucas}},
  \bibinfo{author}{\bibfnamefont{L.}~\bibnamefont{Morellón}},
  \bibinfo{author}{\bibfnamefont{P.~A.} \bibnamefont{Algarabel}},
  \bibinfo{author}{\bibfnamefont{J.~M.} \bibnamefont{De~Teresa}},
  \bibnamefont{and} \bibinfo{author}{\bibfnamefont{M.~R.}
  \bibnamefont{Ibarra}}, \bibinfo{journal}{Nano Letters}
  \textbf{\bibinfo{volume}{15}}, \bibinfo{pages}{492} (\bibinfo{year}{2015}).

\bibitem[{\citenamefont{Aso et~al.}(2014)\citenamefont{Aso, Kan, Shimakawa, and
  Kurata}}]{AFMSRO}
\bibinfo{author}{\bibfnamefont{R.}~\bibnamefont{Aso}},
  \bibinfo{author}{\bibfnamefont{D.}~\bibnamefont{Kan}},
  \bibinfo{author}{\bibfnamefont{Y.}~\bibnamefont{Shimakawa}},
  \bibnamefont{and} \bibinfo{author}{\bibfnamefont{H.}~\bibnamefont{Kurata}},
  \bibinfo{journal}{Advanced Functional Materials}
  \textbf{\bibinfo{volume}{24}}, \bibinfo{pages}{5177} (\bibinfo{year}{2014}),
  ISSN \bibinfo{issn}{1616-3028},
  \urlprefix\url{http://dx.doi.org/10.1002/adfm.201303521}.

\bibitem[{\citenamefont{Aruta et~al.}(2009)\citenamefont{Aruta, Ghiringhelli,
  Bisogni, Braicovich, Brookes, Tebano, and Balestrino}}]{PRBAruta}
\bibinfo{author}{\bibfnamefont{C.}~\bibnamefont{Aruta}},
  \bibinfo{author}{\bibfnamefont{G.}~\bibnamefont{Ghiringhelli}},
  \bibinfo{author}{\bibfnamefont{V.}~\bibnamefont{Bisogni}},
  \bibinfo{author}{\bibfnamefont{L.}~\bibnamefont{Braicovich}},
  \bibinfo{author}{\bibfnamefont{N.~B.} \bibnamefont{Brookes}},
  \bibinfo{author}{\bibfnamefont{A.}~\bibnamefont{Tebano}}, \bibnamefont{and}
  \bibinfo{author}{\bibfnamefont{G.}~\bibnamefont{Balestrino}},
  \bibinfo{journal}{Phys. Rev. B} \textbf{\bibinfo{volume}{80}},
  \bibinfo{pages}{014431} (\bibinfo{year}{2009}),
  \urlprefix\url{http://link.aps.org/doi/10.1103/PhysRevB.80.014431}.

\bibitem[{\citenamefont{Xiaofang et~al.}(2014)\citenamefont{Xiaofang, Long,
  Yang, Christian~M., Shuai, Hui, Xiaoqiang, Shengqi, Lirong, Jing
  et~al.}}]{NatCommZhai}
\bibinfo{author}{\bibfnamefont{Z.}~\bibnamefont{Xiaofang}},
  \bibinfo{author}{\bibfnamefont{C.}~\bibnamefont{Long}},
  \bibinfo{author}{\bibfnamefont{L.}~\bibnamefont{Yang}},
  \bibinfo{author}{\bibfnamefont{S.}~\bibnamefont{Christian~M.}},
  \bibinfo{author}{\bibfnamefont{D.}~\bibnamefont{Shuai}},
  \bibinfo{author}{\bibfnamefont{L.}~\bibnamefont{Hui}},
  \bibinfo{author}{\bibfnamefont{Z.}~\bibnamefont{Xiaoqiang}},
  \bibinfo{author}{\bibfnamefont{C.}~\bibnamefont{Shengqi}},
  \bibinfo{author}{\bibfnamefont{Z.}~\bibnamefont{Lirong}},
  \bibinfo{author}{\bibfnamefont{Z.}~\bibnamefont{Jing}}, \bibnamefont{et~al.},
  \bibinfo{journal}{Nature Communications} \textbf{\bibinfo{volume}{5}},
  \bibinfo{pages}{4253} (\bibinfo{year}{2014}),
  \urlprefix\url{http://www.nature.com/ncomms/2014/140709/ncomms5283/full/ncomms5283.html}.

\bibitem[{\citenamefont{Suntivich et~al.}(2011)\citenamefont{Suntivich,
  Gasteiger, Yabuuchi, Nakanishi, Goodenough, and
  Shao-Horn}}]{NatChemGoodenough}
\bibinfo{author}{\bibfnamefont{J.}~\bibnamefont{Suntivich}},
  \bibinfo{author}{\bibfnamefont{H.~A.} \bibnamefont{Gasteiger}},
  \bibinfo{author}{\bibfnamefont{N.}~\bibnamefont{Yabuuchi}},
  \bibinfo{author}{\bibfnamefont{H.}~\bibnamefont{Nakanishi}},
  \bibinfo{author}{\bibfnamefont{J.~B.} \bibnamefont{Goodenough}},
  \bibnamefont{and}
  \bibinfo{author}{\bibfnamefont{Y.}~\bibnamefont{Shao-Horn}},
  \bibinfo{journal}{Nature Chemistry} \textbf{\bibinfo{volume}{3}},
  \bibinfo{pages}{546} (\bibinfo{year}{2011}).

\bibitem[{\citenamefont{Doennig and Pentcheva}(2015)}]{Pentcheva}
\bibinfo{author}{\bibfnamefont{D.}~\bibnamefont{Doennig}} \bibnamefont{and}
  \bibinfo{author}{\bibfnamefont{R.}~\bibnamefont{Pentcheva}},
  \bibinfo{journal}{Scientific Reports} \textbf{\bibinfo{volume}{5}},
  \bibinfo{pages}{7909} (\bibinfo{year}{2015}).

\bibitem[{\citenamefont{Boschker
  et~al.}(2011{\natexlab{a}})\citenamefont{Boschker, Huijben, Vailionis,
  Verbeeck, van Aert, Luysberg, Bals, van Tendeloo, Houwman, Koster
  et~al.}}]{Koster2011}
\bibinfo{author}{\bibfnamefont{H.}~\bibnamefont{Boschker}},
  \bibinfo{author}{\bibfnamefont{M.}~\bibnamefont{Huijben}},
  \bibinfo{author}{\bibfnamefont{A.}~\bibnamefont{Vailionis}},
  \bibinfo{author}{\bibfnamefont{J.}~\bibnamefont{Verbeeck}},
  \bibinfo{author}{\bibfnamefont{S.}~\bibnamefont{van Aert}},
  \bibinfo{author}{\bibfnamefont{M.}~\bibnamefont{Luysberg}},
  \bibinfo{author}{\bibfnamefont{S.}~\bibnamefont{Bals}},
  \bibinfo{author}{\bibfnamefont{G.}~\bibnamefont{van Tendeloo}},
  \bibinfo{author}{\bibfnamefont{E.~P.} \bibnamefont{Houwman}},
  \bibinfo{author}{\bibfnamefont{G.}~\bibnamefont{Koster}},
  \bibnamefont{et~al.}, \bibinfo{journal}{Journal of Physics D: Applied
  Physics} \textbf{\bibinfo{volume}{44}}, \bibinfo{pages}{205001}
  (\bibinfo{year}{2011}{\natexlab{a}}),
  \urlprefix\url{http://stacks.iop.org/0022-3727/44/i=20/a=205001}.

\bibitem[{\citenamefont{Glazer}(1972)}]{Glazer}
\bibinfo{author}{\bibfnamefont{A.~M.} \bibnamefont{Glazer}},
  \bibinfo{journal}{Acta Crystallogr. Sect. B} \textbf{\bibinfo{volume}{28}},
  \bibinfo{pages}{3384} (\bibinfo{year}{1972}).

\bibitem[{\citenamefont{Vailionis et~al.}(2014)\citenamefont{Vailionis,
  Boschker, Liao, Smit, Rijnders, Huijben, and Koster}}]{Vailionis2014}
\bibinfo{author}{\bibfnamefont{A.}~\bibnamefont{Vailionis}},
  \bibinfo{author}{\bibfnamefont{H.}~\bibnamefont{Boschker}},
  \bibinfo{author}{\bibfnamefont{Z.}~\bibnamefont{Liao}},
  \bibinfo{author}{\bibfnamefont{J.~R.~A.} \bibnamefont{Smit}},
  \bibinfo{author}{\bibfnamefont{G.}~\bibnamefont{Rijnders}},
  \bibinfo{author}{\bibfnamefont{M.}~\bibnamefont{Huijben}}, \bibnamefont{and}
  \bibinfo{author}{\bibfnamefont{G.}~\bibnamefont{Koster}},
  \bibinfo{journal}{Applied Physics Letters} \textbf{\bibinfo{volume}{105}},
  \bibinfo{eid}{131906} (\bibinfo{year}{2014}),
  \urlprefix\url{http://scitation.aip.org/content/aip/journal/apl/105/13/10.1063/1.4896969}.

\bibitem[{\citenamefont{Vila-Fungueiriño
  et~al.}(2015)\citenamefont{Vila-Fungueiriño, Rivas-Murias,
  Rodríguez-González, Txoperena, Ciudad, Hueso, Lazzari, and
  Rivadulla}}]{ACSJose2015}
\bibinfo{author}{\bibfnamefont{J.~M.} \bibnamefont{Vila-Fungueiriño}},
  \bibinfo{author}{\bibfnamefont{B.}~\bibnamefont{Rivas-Murias}},
  \bibinfo{author}{\bibfnamefont{B.}~\bibnamefont{Rodríguez-González}},
  \bibinfo{author}{\bibfnamefont{O.}~\bibnamefont{Txoperena}},
  \bibinfo{author}{\bibfnamefont{D.}~\bibnamefont{Ciudad}},
  \bibinfo{author}{\bibfnamefont{L.~E.} \bibnamefont{Hueso}},
  \bibinfo{author}{\bibfnamefont{M.}~\bibnamefont{Lazzari}}, \bibnamefont{and}
  \bibinfo{author}{\bibfnamefont{F.}~\bibnamefont{Rivadulla}},
  \bibinfo{journal}{ACS Applied Materials \& Interfaces}
  \textbf{\bibinfo{volume}{7}}, \bibinfo{pages}{5410} (\bibinfo{year}{2015}).

\bibitem[{\citenamefont{Smit and Beljers}(1955)}]{SmitBeljers}
\bibinfo{author}{\bibfnamefont{J.}~\bibnamefont{Smit}} \bibnamefont{and}
  \bibinfo{author}{\bibfnamefont{H.}~\bibnamefont{Beljers}},
  \bibinfo{journal}{Philips Res. Rep.} \textbf{\bibinfo{volume}{10}},
  \bibinfo{pages}{113} (\bibinfo{year}{1955}).

\bibitem[{\citenamefont{Barturen et~al.}(2015)\citenamefont{Barturen, Milano,
  Vásquez-Mansilla, Helman, Barral, Llois, Eddrief, and
  Marangolo}}]{MilanoPRB92}
\bibinfo{author}{\bibfnamefont{M.}~\bibnamefont{Barturen}},
  \bibinfo{author}{\bibfnamefont{J.}~\bibnamefont{Milano}},
  \bibinfo{author}{\bibfnamefont{M.}~\bibnamefont{Vásquez-Mansilla}},
  \bibinfo{author}{\bibfnamefont{C.}~\bibnamefont{Helman}},
  \bibinfo{author}{\bibfnamefont{M.~A.} \bibnamefont{Barral}},
  \bibinfo{author}{\bibfnamefont{A.~M.} \bibnamefont{Llois}},
  \bibinfo{author}{\bibfnamefont{M.}~\bibnamefont{Eddrief}}, \bibnamefont{and}
  \bibinfo{author}{\bibfnamefont{M.}~\bibnamefont{Marangolo}},
  \bibinfo{journal}{Phys. Rev. B} \textbf{\bibinfo{volume}{92}},
  \bibinfo{pages}{054418} (\bibinfo{year}{2015}),
  \urlprefix\url{http://link.aps.org/doi/10.1103/PhysRevB.92.054418}.

\bibitem[{\citenamefont{Vittoria}(1993)}]{Vittoria}
\bibinfo{author}{\bibfnamefont{C.}~\bibnamefont{Vittoria}},
  \emph{\bibinfo{title}{Microwave properties of magnetic films}}
  (\bibinfo{publisher}{World Scientific}, \bibinfo{year}{1993}), ISBN
  \bibinfo{isbn}{9781118211494}.

\bibitem[{\citenamefont{Steenbeck and Hiergeist}(1999)}]{Steenbeck}
\bibinfo{author}{\bibfnamefont{K.}~\bibnamefont{Steenbeck}} \bibnamefont{and}
  \bibinfo{author}{\bibfnamefont{R.}~\bibnamefont{Hiergeist}},
  \bibinfo{journal}{Applied Physics Letters} \textbf{\bibinfo{volume}{75}},
  \bibinfo{pages}{1778} (\bibinfo{year}{1999}),
  \urlprefix\url{http://scitation.aip.org/content/aip/journal/apl/75/12/10.1063/1.124817}.

\bibitem[{\citenamefont{Tsui et~al.}(2000)\citenamefont{Tsui, Smoak, Nath, and
  Eom}}]{EomAPL}
\bibinfo{author}{\bibfnamefont{F.}~\bibnamefont{Tsui}},
  \bibinfo{author}{\bibfnamefont{M.~C.} \bibnamefont{Smoak}},
  \bibinfo{author}{\bibfnamefont{T.~K.} \bibnamefont{Nath}}, \bibnamefont{and}
  \bibinfo{author}{\bibfnamefont{C.~B.} \bibnamefont{Eom}},
  \bibinfo{journal}{Applied Physics Letters} \textbf{\bibinfo{volume}{76}},
  \bibinfo{pages}{2421} (\bibinfo{year}{2000}),
  \urlprefix\url{http://scitation.aip.org/content/aip/journal/apl/76/17/10.1063/1.126363}.

\bibitem[{\citenamefont{Alejandro et~al.}(2010)\citenamefont{Alejandro,
  Otero-Leal, Granada, Laura-Ccahuana, Tovar, Winkler, and T.}}]{Alejandro}
\bibinfo{author}{\bibfnamefont{G.}~\bibnamefont{Alejandro}},
  \bibinfo{author}{\bibfnamefont{M.}~\bibnamefont{Otero-Leal}},
  \bibinfo{author}{\bibfnamefont{M.}~\bibnamefont{Granada}},
  \bibinfo{author}{\bibfnamefont{D.}~\bibnamefont{Laura-Ccahuana}},
  \bibinfo{author}{\bibfnamefont{M.}~\bibnamefont{Tovar}},
  \bibinfo{author}{\bibfnamefont{E.}~\bibnamefont{Winkler}}, \bibnamefont{and}
  \bibinfo{author}{\bibfnamefont{C.~M.} \bibnamefont{T.}},
  \bibinfo{journal}{Journal of Physics: Condensed Matter}
  \textbf{\bibinfo{volume}{22}}, \bibinfo{pages}{25602} (\bibinfo{year}{2010}).

\bibitem[{\citenamefont{Ivanshin et~al.}(2010)\citenamefont{Ivanshin,
  Deisenhofer, Krug~von Nidda, Loidl, and Balbashov}}]{Ivanshin}
\bibinfo{author}{\bibfnamefont{V.~A.} \bibnamefont{Ivanshin}},
  \bibinfo{author}{\bibfnamefont{J.}~\bibnamefont{Deisenhofer}},
  \bibinfo{author}{\bibfnamefont{H.-A.} \bibnamefont{Krug~von Nidda}},
  \bibinfo{author}{\bibfnamefont{A.~A.} \bibnamefont{Loidl},
  \bibfnamefont{A.and~Mukhin}}, \bibnamefont{and}
  \bibinfo{author}{\bibfnamefont{M.~V.} \bibnamefont{Balbashov},
  \bibfnamefont{A.~M.and~Eremin}}, \bibinfo{journal}{Physical Review B}
  \textbf{\bibinfo{volume}{61}}, \bibinfo{pages}{6213} (\bibinfo{year}{2010}).

\bibitem[{\citenamefont{Boschker
  et~al.}(2011{\natexlab{b}})\citenamefont{Boschker, Huijben, Vailionis,
  Verbeeck, van Aert, Luysberg, Bals, van Tendeloo, Houwman, Koster
  et~al.}}]{Boschker2011}
\bibinfo{author}{\bibfnamefont{H.}~\bibnamefont{Boschker}},
  \bibinfo{author}{\bibfnamefont{M.}~\bibnamefont{Huijben}},
  \bibinfo{author}{\bibfnamefont{A.}~\bibnamefont{Vailionis}},
  \bibinfo{author}{\bibfnamefont{J.}~\bibnamefont{Verbeeck}},
  \bibinfo{author}{\bibfnamefont{S.}~\bibnamefont{van Aert}},
  \bibinfo{author}{\bibfnamefont{M.}~\bibnamefont{Luysberg}},
  \bibinfo{author}{\bibfnamefont{S.}~\bibnamefont{Bals}},
  \bibinfo{author}{\bibfnamefont{G.}~\bibnamefont{van Tendeloo}},
  \bibinfo{author}{\bibfnamefont{E.~P.} \bibnamefont{Houwman}},
  \bibinfo{author}{\bibfnamefont{G.}~\bibnamefont{Koster}},
  \bibnamefont{et~al.}, \bibinfo{journal}{Journal of Magn. Magnetic Mat.}
  \textbf{\bibinfo{volume}{323}}, \bibinfo{pages}{2632}
  (\bibinfo{year}{2011}{\natexlab{b}}).

\bibitem[{\citenamefont{Belmeguenai et~al.}(2010)\citenamefont{Belmeguenai,
  Mercone, Adamo, Méchin, Fur, Monod, Moch, and Schlom}}]{BelmeguenaiPRB2010}
\bibinfo{author}{\bibfnamefont{M.}~\bibnamefont{Belmeguenai}},
  \bibinfo{author}{\bibfnamefont{S.}~\bibnamefont{Mercone}},
  \bibinfo{author}{\bibfnamefont{C.}~\bibnamefont{Adamo}},
  \bibinfo{author}{\bibfnamefont{L.}~\bibnamefont{Méchin}},
  \bibinfo{author}{\bibfnamefont{C.}~\bibnamefont{Fur}},
  \bibinfo{author}{\bibfnamefont{P.}~\bibnamefont{Monod}},
  \bibinfo{author}{\bibfnamefont{P.}~\bibnamefont{Moch}}, \bibnamefont{and}
  \bibinfo{author}{\bibfnamefont{D.~G.} \bibnamefont{Schlom}},
  \bibinfo{journal}{Phys. Rev. B} \textbf{\bibinfo{volume}{81}},
  \bibinfo{pages}{054410} (\bibinfo{year}{2010}).

\bibitem[{\citenamefont{Cullity and Graham}(2011)}]{cullity}
\bibinfo{author}{\bibfnamefont{B.}~\bibnamefont{Cullity}} \bibnamefont{and}
  \bibinfo{author}{\bibfnamefont{C.}~\bibnamefont{Graham}},
  \emph{\bibinfo{title}{Introduction to Magnetic Materials}}
  (\bibinfo{publisher}{Wiley}, \bibinfo{year}{2011}), ISBN
  \bibinfo{isbn}{9781118211496}.

\bibitem[{\citenamefont{Ikeda et~al.}(2010)\citenamefont{Ikeda, Miura,
  Yamamoto, Mizunuma, Gan, Endo, Kanai, Hayakawa, Matsukura, and
  Ohno}}]{Ikeda2010}
\bibinfo{author}{\bibfnamefont{S.}~\bibnamefont{Ikeda}},
  \bibinfo{author}{\bibfnamefont{K.}~\bibnamefont{Miura}},
  \bibinfo{author}{\bibfnamefont{H.}~\bibnamefont{Yamamoto}},
  \bibinfo{author}{\bibfnamefont{K.}~\bibnamefont{Mizunuma}},
  \bibinfo{author}{\bibfnamefont{H.}~\bibnamefont{Gan}},
  \bibinfo{author}{\bibfnamefont{M.}~\bibnamefont{Endo}},
  \bibinfo{author}{\bibfnamefont{S.}~\bibnamefont{Kanai}},
  \bibinfo{author}{\bibfnamefont{J.}~\bibnamefont{Hayakawa}},
  \bibinfo{author}{\bibfnamefont{F.}~\bibnamefont{Matsukura}},
  \bibnamefont{and} \bibinfo{author}{\bibfnamefont{H.}~\bibnamefont{Ohno}},
  \bibinfo{journal}{Nature Materials} \textbf{\bibinfo{volume}{9}},
  \bibinfo{pages}{721} (\bibinfo{year}{2010}).

\bibitem[{\citenamefont{Vidal et~al.}(2012)\citenamefont{Vidal, Zheng, Schio,
  Bonilla, Barturen, Milano, Demaille, Fonda, De~Oliveira, and
  Etgens}}]{Vidal2012}
\bibinfo{author}{\bibfnamefont{F.}~\bibnamefont{Vidal}},
  \bibinfo{author}{\bibfnamefont{Y.}~\bibnamefont{Zheng}},
  \bibinfo{author}{\bibfnamefont{P.}~\bibnamefont{Schio}},
  \bibinfo{author}{\bibfnamefont{F.}~\bibnamefont{Bonilla}},
  \bibinfo{author}{\bibfnamefont{M.}~\bibnamefont{Barturen}},
  \bibinfo{author}{\bibfnamefont{J.}~\bibnamefont{Milano}},
  \bibinfo{author}{\bibfnamefont{D.}~\bibnamefont{Demaille}},
  \bibinfo{author}{\bibfnamefont{E.}~\bibnamefont{Fonda}},
  \bibinfo{author}{\bibfnamefont{A.}~\bibnamefont{De~Oliveira}},
  \bibnamefont{and} \bibinfo{author}{\bibfnamefont{V.}~\bibnamefont{Etgens}},
  \bibinfo{journal}{Physical Review Letters} \textbf{\bibinfo{volume}{109}},
  \bibinfo{pages}{117205} (\bibinfo{year}{2012}).

\bibitem[{\citenamefont{Sallica~Leva et~al.}(2010)\citenamefont{Sallica~Leva,
  Valente, Marttínez~Tabares, Vásquez~Mansilla, Roshdestwensky, and
  Butera}}]{Butera2010}
\bibinfo{author}{\bibfnamefont{E.}~\bibnamefont{Sallica~Leva}},
  \bibinfo{author}{\bibfnamefont{R.}~\bibnamefont{Valente}},
  \bibinfo{author}{\bibfnamefont{F.}~\bibnamefont{Marttínez~Tabares}},
  \bibinfo{author}{\bibfnamefont{M.}~\bibnamefont{Vásquez~Mansilla}},
  \bibinfo{author}{\bibfnamefont{S.}~\bibnamefont{Roshdestwensky}},
  \bibnamefont{and} \bibinfo{author}{\bibfnamefont{A.}~\bibnamefont{Butera}},
  \bibinfo{journal}{Physical Review B} \textbf{\bibinfo{volume}{82}},
  \bibinfo{pages}{144410} (\bibinfo{year}{2010}).

\bibitem[{\citenamefont{Kronm\"{u}ller and F\"{a}hnle}(2003)}]{Kronmuller}
\bibinfo{author}{\bibfnamefont{H.}~\bibnamefont{Kronm\"{u}ller}}
  \bibnamefont{and}
  \bibinfo{author}{\bibfnamefont{M.}~\bibnamefont{F\"{a}hnle}},
  \emph{\bibinfo{title}{Micromagnetism and the microstructure of ferromagnetic
  solids}} (\bibinfo{publisher}{Cambridge University Press},
  \bibinfo{year}{2003}), ISBN \bibinfo{isbn}{9781118211495}.

\end{thebibliography}

\end{document}